# Time-temperature superposition in grain and grain boundary response regime of $A_2HoRuO_6$ (A = Ba, Sr, Ca) double perovskite ceramics: A Conductivity Spectroscopic Analysis

Saswata Halder[a*], Alo Dutta[b], T. P. Sinha[a]

[a]*Department of Physics, Bose Institute, 93/1, Acharya Prafulla Chandra Road, Kolkata 700009, India.*

[b]*Department of Condensed Matter Physics and Material Sciences, S. N. Bose National Centre for Basic Sciences, Block-JD, Sector-III, Salt Lake, Kolkata 700098, India.*

Abstract:

The pursuit for an appropriate universal scaling factor to satisfy the time-temperature superposition principle for grain and grain boundary responses has been explored in the ac conductivity domain for polycrystalline double perovskite oxides $A_2HoRuO_6$ (AHR; A = Ba, Sr, Ca). The samples show different structural phases ranging from cubic to monoclinic with decreasing ionic radii. The degree of distortion in the materials is correlated to the strength of bonding through the bond valence sum (BVS). The conductivity spectra for all the samples obey the power law behaviour. The contribution of different microstructural features to the conduction process is established. Thermal variation of dc resistivity points towards a gradual crossover from nearest neighbour to variable range hopping. The activation energies obtained from dc conductivity, hopping frequency and relaxation frequency show close correlation between the conduction and relaxation mechanisms. The scaled conductivity curves for AHR showed the presence of two different conduction processes with dissimilar activation energies in the grain boundary and grain response regimes. It is thus concluded that a single scaling parameter is insufficient to satisfy the time temperature superposition principle universally when two different thermally activated regions are present simultaneously in the materials.

## 1. Introduction

The solid state chemistry of 4d Ru-based materials has always been an intriguing subject to the scientific community since the discovery of spin-glass behavior in $Sr_2FeRuO_6$ [1-3]. The Ru-based double perovskite oxides (DPO) have emerged as a fertile ground for the exploration of novel transport phenomena as it strikes a delicate balance between localised and delocalised electron behaviour driven by spin-orbit coupling (SOC), Coulombic interactions (U) and crystal field interactions [4-15]. In the case of DPO with Ru at the B-sites, an increase in orbital overlap with decrease in unit cell volume has been shown to increase the Ru-d electron delocalisation [16]. The sensitivity of the narrow itinerant 4d band of Ru is dependent on the degree of hybridization between the Ru-$t_{2g}$ and the O-2p orbitals alongwith the distortions present in the octahedral environment of Ru. Since the B-site of a DPO provides a dominant influence in determining the transport properties of the material, Ru at the octahedral B-site plays a major role in determining the electrical and magnetic properties of the host material. This research article focusses on the octahedral distortion induced electrical response of a specific DPO series $A_2HoRuO_6$ (AHR; A = Ba, Sr, Ca) in the radio frequency regime. The crystal and magnetic properties of AHR [17-19] has been studied in details by many research groups but the knowledge of the ac electrical properties in the radio frequency region remains a scarcity in literature.

An understanding of the charge carrier dynamics is imperative in order to tailor the electrical properties of materials. The charge carrier dynamics can be extracted with the help of either the conductivity spectra [20] or the modulus spectra [21]. The conductivity spectra follow a power law behaviour [22, 23] consisting of a frequency independent (dc) region and a frequency dependent dispersive component. The crossover frequency which is also known as the hopping frequency provides the transition from the dc to the ac region. The hopping frequency can be related to the dc conductivity through the Nernst-Einstein relation (Eq. 1) and can be used to estimate the thermal variation of charge carrier concentration

$$\sigma_{dc} = en_c\mu = \frac{n_c e^2 \gamma \lambda^2}{kT} \omega_H \qquad (1)$$

where $n_c$ is the concentration of the mobile charges, μ is their mobility, e is the electronic charge, γ is a geometrical factor related to hopping and λ is the hopping distance. The scaling behaviour of the conductivity spectra of glasses and amorphous materials has been studied previously [24, 25]. But very few studies exist that explores the dynamics of charge carriers in polycrystalline materials. Saha et. al. found the validity of time-temperature superposition (TTSP) for polycrystalline $BaFe_{0.5}Nb_{0.5}O_3$ at low temperatures through the scaling of the ac conductivity spectra [26]. TTSP is satisfied by the scaling behaviour of the ac conductivity spectra when the conductivity isotherm collapses into a single master curve with respect to specified scaling parameters [27] showing the temperature independence of the conduction mechanism. The temperature dependence is incorporated in the increase in the charge carrier density without affecting the conduction mechanism. However, the validation of TTSP at the different microstructural region for polycrystalline samples is rarely studied. The TTSP can be mathematically expressed as [28, 29],

$$\frac{\sigma'(\omega)}{\sigma_{dc}} = F\left(\frac{\omega}{\omega_s}\right) \qquad (2)$$

where the scaling factor F is independent of temperature and $\omega_s$ is the temperature dependent scaling parameter which depends on the various scaling models proposed by different researchers [30-32]. The charge carrier dynamics of AHR is comprehensively discussed in the measured temperature range in the realms of ac conductivity and electric modulus formalisms. The validity of the TTSP is investigated in the various microstructural response regions through the scaling of the ac conductivity spectra in the domain of Summerfield [33] and Ghosh scaling formalisms [34, 35].

## 2. Experimental details: Material synthesis and characterization

AHR was synthesized using the standard solid state reaction method. Powders of $BaCO_3$ (Loba Chemie, 99% pure), $SrCO_3$ (Loba Chemie, 99% pure), $CaCO_3$ (Loba Chemie, 99% pure), $Ho_2O_3$ (Sigma-Aldrich, 99% pure) and $RuO_2$ (Alfa-Aesar, 99% pure) were taken in appropriate proportions in accordance with stoichiometric analysis. The ingredients were mixed homogeneously by grinding in dry form and then in acetone medium for 10 h. The mixture was calcined in an alumina crucible at 1300 ºC for 10 h and brought to room temperature at the cooling rate of 90° C/h. In the next step, the calcined powder was pelletized into discs of 8 mm diameter and 1.7 mm thickness using polyvinyl alcohol as binder. Finally, the discs were sintered at 1350 ºC and cooled down to room temperature at the rate of 1.6 ºC/min.

The X-ray diffraction (XRD) patterns of the samples were collected using X-ray powder diffractometer (Rigaku Miniflex II diffractometer, Tokyo, Japan) using CuK$\alpha$ radiation in the 2θ range of 10º - 80º at a scanning rate of 0.02º per step. Rietveld refinement of the XRD data is performed by the Fullprof code [36]. For dielectric measurements, the sintered pellets were polished on both the sides and kept between two thin gold electrodes. The complex impedance (Z*), phase angle (ϕ) and conductance (G) were measured using a computer controlled LCR meter (HIOKI-3552, Nagano, Japan) in the frequency range from 42 Hz to 5 MHz and temperature range 303 K to 613 K at an oscillation voltage of 1.0 V. The temperature was controlled by a Eurotherm 2216e programmable temperature controller connected with the oven and each measured temperature was kept constant with an accuracy of ±1 K.

3. Results and Discussions:

3.1. Structural Characterization:

The room temperature XRD patterns of BHR, SHR and CHR are shown in Figs. 1-3 respectively. The structural and fitted parameters for BHR, SHR and CHR are given in Table 1. The Rietveld refinement of the XRD pattern of BHR shows a cubic phase with space group *Fm3m*. The space group allows a 1:1 ordered arrangement between $Ho^{3+}$ and $Ru^{5+}$ ions. Superlattice lines such as [111] and [311] which arise from the alternate ordering of Ho and Ru sites in BHR are observed in the XRD pattern (Fig. 1). The intensity ratio of the superlattice reflection peaks I[111]/I[220] measures the degree of site ordering in the cubic lattice. The intensity ratio is found to be 0.04, indicating the ordering in the Ho and Ru site in BHR.

The XRD profiles of SHR and CHR with monoclinic space group *P2$_1$/n* present contrasting features. The XRD pattern of SHR sample in Fig. 2 show two superlattice reflection peaks at $2\theta \approx 18°$ [$\bar{1}$01] and 24° [$\bar{1}$11] respectively. The degree of site ordering in the monoclinic lattice is given by the intensity ratio $I[\bar{1}01]/I[\bar{1}12]$ of the superlattice reflection peak. This intensity ratio is found to be 0.03, indicating the ordering of Ho and Ru at the B-site in SHR. The reflection at $2\theta \approx 18°$ [$\bar{1}$01] is a consequence of the antiphase tilting of the octahedra around the principal 2-fold axis [37]. The other reflection located at $2\theta \approx 24°$[$\bar{1}$11], arises from the in-phase tilt of the octahedra around the primitive axis [38]. The total intensity of the [002] and [110] reflections are significantly small when Ca replaces Sr at the A-site as evident from the comparative study of the XRD profiles of SHR and CHR thus showing the A-site is partially occupied by the $Ca^{2+}$ ions giving rise to a different cationic distribution in CHR as compared to SHR. The cationic distributions that best fits the intensity profile of CHR is $Ca^{2+}/Ho^{3+}$ (4e), $Ca^{2+}/Ho^{3+}$ (2d), $Ru^{5+}$ (2c) corresponding to the formula $(Ca_{1.46}Ho_{0.54})(Ca_{0.54}Ho_{0.46})RuO_6$ showing the partial occupation of the Ca ions and Ho ions at both the A-site and B-site of the double perovskite in agreement to the suggestion by Battle et. al. [19]. There exists no cationic ordering between the $Ca^{2+}$ and $Ho^{3+}$ ions at either of the A or B-sites although the $Ca^{2+}/Ho^{3+}$ ions form a 1:1 ordered arrangement with the $Ru^{5+}$ ions at the B-site. The intensity ratio of the superlattice reflection in CHR ($I[011]/I[\bar{1}12]$) is found to be 0.12 which also signifies the partial B-site ordering in the sample.

Bond valence sum (BVS) and bond distances for AHR are given in Table 1. The BVS of the $A^{2+}$ cations are significantly less than 2.0 which demonstrates that the average A-O distances are longer than optimal [39]. The BVS for $Ba^{2+}$ and $Sr^{2+}$ in BHR and SHR is 1.992 and 1.783 respectively. The $Ca^{2+}$ (BVS = 1.382) at the A-site is comparatively more under-bonded than the $Ca^{2+}$ (BVS = 1.801) at the B-site showing that the bonding strength in $CaO_6$ octahedra is more than in the $CaO_{12}$ octahedra. The increased bond valency reflects that the A-site is becoming more convenient for the $A^{2+}$ cation thereby reducing the requirement for octahedral tilting resulting in a more symmetrical crystal structure. The bond lengths and BVS provide further insight into the bonding requirement of the two B-site cations of significant size difference which are distributed in an ordered perovskite structure. In case of BHR and SHR, the $Ho^{3+}$ are over-bonded with the Ho-O bond lengths being shorter than the expected value. In the case of CHR, both Ho atoms are considerably under-bonded which is reflected in the respective bond lengths which are greater than the calculated values. This can be attributed to the increase in the distortion resulting in the lengthening and shortening of different axes to attain the lowest energy stable crystal structure. The Ru-atom in all three materials are overbonded as observed from Ru-O bond length is shorter than the expected value. The Ho-O bond lengths systematically increase whereas the Ru-O bond length decreases as Ca replaces Ba/Sr at the A-site. This shows that with distortion the Ho-O axis elongates and Ru-O axis shortens. The suitability of any cation pair to form a perovskite structure is expressed in terms of the tolerance factor [40, 41]. It predicts that for a given pair of B-site cations an increase in the size of the A-site cation reduces the distortion in the perovskite system. The values of the geometric tolerance factor $T_f$ for BHR, SHR and CHR are calculated to be 1.004, 0.95 and 0.91 respectively, which is in agreement with the cubic structure of BHR and the monoclinic structures for SHR and CHR respectively as shown in the insets of Figs. 1-3. The increased bond valency reflects that the size of the A-site is becoming more suitable for the $A^{2+}$ cation reducing the requirement for octahedral tilting in the perovskite structure resulting in an increase in the symmetry of the materials due to the volumetric resemblance of the $BO_6$ and $AO_{12}$ octahedra. The crystal structure approaches the higher symmetry configuration as one moves from CHR to BHR reducing the octahedral distortion in the system as is reflected in the <B′-O-B″> bond angles which approach 180°.

3.2. Electrical Characterization:

Dynamical processes in materials are studied by a variety of spectroscopic techniques. Electrical relaxation measurements are commonly used, mostly as a function of frequency to study the charge dynamics in materials [42]. With the advancement of ac conductivity measurement techniques, impedance analysis has become a widely accepted tool for the characterization of the experimental data. Figs. 4(a)-(c) present the spectroscopic plots of the real part of ac conductivity for BHR, SHR and CHR respectively at selected temperatures. The symbols represent the experimental data and the solid lines represent the fit of the experimental conductivity spectra in accordance to the power (Eq. (3)) suggested by Almond and West: [43, 44]

$$\sigma'(\omega) = \sigma_{dc}\left[1 + \left(\frac{\omega}{\omega_H}\right)^n\right] \quad (3)$$

In Eq. (3), $\sigma_{dc}$ represents the dc conductivity, $\omega$ is the angular frequency, n is the power law exponent describing the electrical relaxation behaviour of the material and $\omega_H$ represents the hopping frequency of the carriers which marks the crossover from dc to the dispersive conduction region at $\omega > \omega_H$. In a polycrystalline ceramic system, conductivity relaxation occurs due to grain, grain-boundary and electrode-specimen interface contributions. In absence of electrode polarization, electro-ceramics show the presence of plateaus and dispersion regions in their conductivity spectra due to the contribution from different microstructural regions [45-47]. The spectroscopic plots for BHR, SHR and CHR have been divided into different regimes revealing different microstructural contribution to the conductivity of the materials. In the case of BHR (Fig. 4(a)), a single plateau in the low frequency region (Region I) followed by a dispersion region (Region II) is observed. The single plateau can be assigned to the total conductivity of the sample measuring the overall dc conductivity of the sample in the limit $\omega \to 0$ i.e. ($\sigma_{dc} = \sigma_t$ ($\omega \to 0$)). The dispersion region can be attributed to the grain boundary relaxation of the sample. In the case of SHR and CHR (Fig. 4(b)-(c)), the presence of another plateau is observed along with Regions I and II. This plateau denoted by Region III can be assigned to the contribution of the grains to the total conductivity.

The frequency dependent conductivity can be interpreted in terms of the grain boundary conductivity and jump relaxation model [48, 49]. In this model, the total conductivity, which is a measure of the dc conductivity, is caused by the translational motion of charge carriers between localised sites. The ac conductivity in the dispersion regime is caused by the correlated backward and forward hopping motion of the charge carriers which can be visualised as a competition between two relaxation processes; successful and unsuccessful hopping of charge carriers. The conductivity of BHR is found to be greater than SHR and CHR for a particular temperature. The conductivity of the materials are enhanced with increase in temperature due to the thermal activation of charge carriers across the energy gap. As previously mentioned the conduction in AHR depends on the degree of hybridization between the Ru $t_{2g}$ and O-2$p$ orbitals along with the different octahedral distortions present in the Ru surroundings. As Ca replaces Ba and Sr at the A-site of the perovskite lattice, the degree of distortion increase which lowers the hybridization between the Ru $t_{2g}$ and O-2$p$ orbitals. The decrease in the delocalisation of electrons around the Ru-atom inhibits the facilitative transfer of electrons from the valence band to the conduction band. Thus the conductivity decreases as the distortions increase from BHR (T$_f$ = 1.004) to CHR (T$_f$ = 0.91). It seems plausible that due to the ordered arrangement of the octahedra the complete delocalization of Ru cation could be inhibited by the HoO6 octahedra by physical separation of the two successive RuO6 octahedra. In AHR, every second Ru is substituted by Ho forming a rock-salt arrangement and disrupting the Ru-O perovskite motif of intersecting 1D chains. As a result the AHR shows more localised electrical behaviour than most of the parent perovskite derived ruthenates [50-54].

If the dc conductivity is Arrhenius activated, then $\sigma_t \propto \frac{E_a^\sigma}{kT}$, where E$^\sigma_a$ is the activation energy for dc conductivity. For a fixed number of sites N and fixed hopping length R, it can be shown $\omega_H \propto \frac{E_a^\omega}{kT}$ [27]. If the power law exponent '$n$' is independent or a weakly varying function of temperature then $\omega_H$ activates with the same energy as $\sigma_t$. In the present study, '$n$' is found to be almost independent of temperature. Arrhenius representation of dc conductivity and hopping rate for BHR, SHR and CHR are shown in Figs. 5 and 6 respectively. The activation energy for grain boundary is higher than the activation energy for grains. Both dc conductivity ($\sigma_t$) and hopping frequency ($\omega_H$) show identical profiles with a linear behaviour in the investigated temperature range. The activation energies for $\sigma_t$ and $\omega_H$, determined from the slope of the linear fit to the experimental data, are given in Table 2. It is observed that the activation energies of dc conductivity are close to the activation energies of hopping frequency for all the three samples showing that $\sigma_t$ is proportional to $\omega_H$. The value of the activation energy for bulk conductivity suggests that the conduction mechanism occurs due to the hopping of small polarons. In general, an activation energy of more than 0.2 eV is reminiscent of polaronic conduction of holes. On this basis it can be inferred that p-type carriers are responsible for the charge conduction in AHR [55, 56].

To examine the correlation between dc conductivity ($\sigma_t$) and hopping frequency ($\omega_H$), the logarithmic plot were drawn between these two parameters as shown in Figs. 7(a)-(c). The figures show a near linear dependence of these parameters on each other with a slope of almost unity indicating a power law dependence of $\sigma_{dc}$ on $\omega_H$ of the form $\sigma_t \propto \omega_H^n$ where n is the slope of the graph. This implies that the dc and ac conductivity are correlated in AHR [57].

The electrical resistivity $\rho$ can be deduced from the dc conductivity $\sigma_t$ using the relation $\rho = 1/\sigma_t$. To understand the possible conduction mechanism, it is customary to investigate the variation of electrical resistivity with temperature. The results are analysed in terms of the hopping models namely, (i) small polaron hopping (SPH) or nearest neighbour hopping and (ii) variable range hopping (VRH) [58, 59]. The general form of the hopping equation can be written as $\rho \propto e^{\left(\frac{A}{T}\right)^x}$ where A = $T_o$ and $x = 0.25$ represents VRH whereas A = $E_a/k_B$ and $x = 1$ describes the nearest neighbour hopping limit. In order to obtain the exponent '$x$' which determines the nature of the hopping process, a plot of (log log $\rho$) vs log T is shown in Figs. 8(a)-(c) for BHR, SHR and CHR respectively. The slope of the graph determines the value of $x$ [60]. In all the three materials, the high temperature domain is dominated by the nearest neighbour hopping process ($x \sim 1$) with a gradual cross-over towards variable range hopping limit near room temperature ($x \sim 0.30$). With further decrease in temperature, the VRH limit of $x \sim 0.25$ should be achieved.

To gain valuable insight into the temperature dependence of conduction mechanism and the response of grains and grain boundaries to the conductivity, the scaling behaviour of frequency dependent conductivity has been performed. In the present study two different scaling formalisms, Summerfield scaling [33] and Ghosh scaling [34, 35], have been applied to perform the scaling of the conductivity spectroscopic plots. The conductivity axis is scaled by the dc conductivity ($\sigma_t$) whereas the frequency axis is scaled by $\omega_s = \sigma_t T$ and $\omega_s = \omega_H$ representing the Summerfield and the Ghosh scaling respectively. Figs. 9(a)-(c) and 10(a)-(c) represent the Summerfield and Ghosh scaling of the ac conductivity spectra for BHR, SHR and CHR respectively. The ac conductivity spectra for BHR, SHR and CHR are superimposed into a single master curve at the grain boundary region although in the grain response region the ac conductivity data do not collapse into a single master curve. It shows that the relaxation processes at the grain boundary region follow the time-temperature superposition principle (TTSP). However, the deviation of the scaled conductivity curves at the grain response region implies that the relaxation phenomena in the grains are different from that in the associated grain boundary response regions. The grain boundaries require higher activation energy than the grains as is evident from Table 2. This difference is in agreement with the dissimilar activation energies of the grains and the grain boundaries ($\Delta E = E_a^t - E_a^g = 0.09$ eV for SHR and 0.11 for CHR). The higher activation energy of the grain boundary conductivity can be due to the combination of a number of serial barriers in the interfacial region and therefore the total conductivity is dictated by the grain boundary impedance. The parameters, such as the diffusion coefficient (D), which affect the bulk and grain conductivity can be obtained from the differently activated response regions. In order to consider the activation energy difference between the bulk and grain interiors the frequency axis has been scaled by the parameter $\frac{\sigma_t T}{e^{\frac{-(E_a^t - E_a^g)}{kT}}}$ and $\frac{\omega_H}{e^{\frac{-(E_a^t - E_a^g)}{kT}}}$ [61] for SHR as shown in Figs. 9(d) and 10(d) respectively. The scaled grain boundary curves are superimposed into a single curve whereas the grain response curve deviate more than previously. The factor $e^{\frac{-(E_a^t - E_a^g)}{kT}}$ compensates for the different energy barriers that the mobile ion encounters between the grain boundaries and the grain regions during its percolation. The conductivity scaled isotherms for BHR, SHR and CHR show that the Ghosh scaling formalism ($\omega_s = \omega_H$) provides a better superposition than the Summerfield scaling formalism ($\omega_s = \sigma_t T$) due to the fact that $\sigma_t$ has a power law dependence on $\omega_H$.

Previous studies have revealed that some school of thought prefer the use of electric modulus formalism in the interpretation of the conduction dynamics by eliminating the electrode polarization effect. The electric modulus formalism was first introduced by Macedo et. al. [62] in exploring the space-charge relaxation phenomena and is used in conjunction with the impedance and permittivity formalisms to differentiate between the grain and grain boundary regions and to differentiate between the microscopic processes responsible for localized dielectric relaxations and long-range conductions [63, 64]. Figs. 11(a)-(c) display the spectroscopic plots for the imaginary component of electric modulus, M″, at selected temperatures for BHR, SHR and CHR respectively. BHR displays a single relaxation peak in the entire frequency domain whereas both SHR and CHR have two relaxation peaks, one at low frequency due to grain boundary response and the other at high frequency due to the grain response [65]. The M″ peaks are much broader than an ideal Debye peak showing that the relaxation behaviour is non-ideal and the relaxation times are distributed. The increase in frequency as one approaches the peak from the left facilitates the mobility of the charge carriers and represents a region of long range conduction. To the right of the peak, localized forward backward hopping motion occurs as the charge carriers are confined to trapped centres due to their decreased mobility [66, 67]. With increase in temperature charge carrier dynamics is boosted with a decrease in relaxation time which manifests itself as a shift in the relaxation frequency ($\omega_m$) towards the higher frequency side. A natural conclusion thus follows that the relaxation process is thermally activated with charge carrier hopping determining the conduction dynamics in AHR.

The scaling behaviour of $M''$ has been performed to obtain an insight into the relaxation dynamics of AHR [68]. The insets of Figs. 11(a)-(c) provide the scaled spectra in which each axis has been scaled by its corresponding peak value obtained from the relaxation maxima at individual temperatures. The figures show that the modulus spectra at different temperatures nearly collapse into a single master curve at the grain boundary response regime, whereas the grain response domain shows a deviation from the scaling behaviour. Similar nature was also obtained from the conductivity scaling behaviour. This indicates that the dynamical process controlling the conductivity relaxation at the grain boundaries is the same over the entire temperature range although it deviates greatly from the conductivity relaxation dynamics at the grain region. It is therefore concluded that there exists no single scaling parameter that can account for the validation of the time-temperature superposition principle when two dissimilar thermally activated processes are simultaneously present within a material.

Fig. 11(d) represents the thermal variation of relaxation frequency ($\omega_m$) for AHR. The activation energies obtained from the linear fit of the experimental data are 0.25 eV, 0.29 eV and 0.31 eV for BHR, SHR and CHR respectively. A comparison of the activation energies of all the samples for dc conductivity ($\sigma_t$), hopping frequency ($\omega_H$) and relaxation frequency ($\omega_m$) show that the values lie close to each other. The close proximity of the values of $E_H$ and $E_{\omega max}$ indicate that the hopping frequency and the relaxation frequency provide the limiting bound to the crossover from short range hopping to long range percolation.

## 4. Conclusion:

A comprehensive quantitative investigation of electrical conduction for polycrystalline double perovskite oxide ceramics $A_2HoRuO_6$ (AHR; A = Ba, Sr, Ca) prepared by solid state technique has been presented over the temperature range of 303 K-613 K and frequency range 42 Hz-5 MHz. The room temperature XRD data confirmed the single phase formation of all the materials. The degree of octahedral distortion has been correlated to the strength of the bonding of the A-site cation. The conductivity spectra for AHR follow a power law variation. The contribution of different mictrostructural features to the conduction mechanism are observed in the conductivity spectra. The dc conductivity and the hopping frequency are correlated in a power law form. The activation energy of the materials in the grain boundary lies in the range of 0.27-0.37 eV whereas the grain response region have energies of 0.22-0.26 eV. The polaronic conduction mechanism shows a gradual crossover from nearest neighbour hopping to variable range hopping with decrease in temperature. The electric modulus spectra show a thermally activated conduction mechanism. The activation energies obtained from linear fit of the thermal variation of relaxation frequency ($\omega_m$) are in close proximity to the activation energies obtained from the hopping frequencies ($\omega_H$) showing that both ($\omega_m$) and ($\omega_H$) are correlated. Our investigations show that the time temperature superposition principle using both the Summerfield scaling and the Ghosh scaling are valid only at the specific microstructural response domains and not universally. Hence, we conclude if two microstructural regions have simultaneous dissimilar thermally activated conduction processes in them then it is not possible to obtain a single scaling parameter that can account for the TTSP.

Captions to Tables:

Table 1: Crystal structure parameters for AHR as obtained from Rietveld Refinement of XRD data.

Table 2: Bond valence sum (BVS) and Activation energies of Grain and Grain Boundary from Arrhenius Fit of thermal variation of dc conductivity, hopping frequency and relaxation frequency.

Captions to Figures:

Figure 1: Room temperature XRD of $Ba_2HoRuO_6$ (BHR). The inset shows the crystal structure.

Figure 2: Room temperature XRD of $Sr_2HoRuO_6$ (BHR). The inset shows the crystal structure.

Figure 3: Room temperature XRD of $Ca_2HoRuO_6$ (BHR). The inset shows the crystal structure.

Figure 4(a): ac conductivity spectroscopic plot for BHR at selected temperatures.

Figure 4(b): ac conductivity spectroscopic plot for SHR at selected temperatures.

Figure 4(c): ac conductivity spectroscopic plot for CHR at selected temperatures.

Figure 5: Thermal variation of dc conductivity ($\sigma_{dc}$) for (a) BHR (b) SHR (c) CHR.

Figure 6: Thermal variation of hopping frequency ($\omega_H$) at grain boundaries for (a) BHR (b) SHR (c) CHR.

Figure 7: Correlation between dc conductivity ($\sigma_{dc}$) and hopping frequency ($\omega_H$) at grain boundaries for (a) BHR (b) SHR (c) CHR.

Figure 8: Temperature dependence of electrical resistivity (a) BHR (b) SHR (c) CHR.

Figure 9: Summerfield scaling of ac conductivity for (a) BHR (b) SHR (c) CHR (d) Summerfield scaling for SHR with frequency axis scaled by $\frac{\sigma_t T}{e^{\frac{-(E_a^t - E_a^g)}{kT}}}$.

Figure 10: Scaling of ac conductivity according Ghosh formalism for (a) BHR (b) SHR (c) CHR (d) Scaling of ac conductivity for SHR with frequency axis scaled by $\frac{\omega_H}{e^{\frac{-(E_a^t - E_a^g)}{kT}}}$.

Figure 11: Frequency dependent electric modulus (M″) for (a) BHR (b) SHR (c) CHR. The insets show the scaling behaviour. (d) Thermal variation of relaxation frequency ($\omega_m$) for BHR, SHR and CHR.

19